\documentstyle[12 pt]{article}
\begin{document}
\title{ Comment on \lq Relativistic shape invariant potentials \rq   }
\author{Arvind Narayan Vaidya$^{\star}$\\
Instituto de F\'\i sica - Universidade Federal do Rio de Janeiro \\
Caixa Postal 68528 - CEP 21945-970, Rio de
Janeiro, Brazil\\
Rafael de Lima Rodrigues\thanks{Permanent address: Departamento de
Ci\^encias Exatas e da Natureza, Universidade Federal da Para\'\i ba,
Cajazeiras - PB, 58.900-000 - Brazil.}  
\\ Centro Brasileiro de Pesquisas F\'\i sicas (CBPF)\\
Rua Dr. Xavier Sigaud, 150,CEP 22290-180,Rio de Janeiro, RJ, Brazil}

\maketitle
\begin{abstract}
In this comment we point out numerous errors in the paper of Alhaidari cited in the title.
\end{abstract}

PACS numbers : 03.65.Pm 03.65.Ge
 
\date{}
$^{\star}$ {e-mail:vaidya@if.ufrj.br}\nonumber\\

In a recent paper Alhaidari [1] treats the problem of formulating  a relativistic Dirac type equation which can be reduced to solving a Schroedinger equation for shape invariant potentials for the upper component while the lower component can be found once the upper component has been found.The method used is same as one used in his earlier paper [2].
\par
The proposed Hamiltonian is (we use the notation of Bjorken and Drell [3])
\begin{equation}
H= {\mbox{\boldmath $\alpha$}}\cdot({\bf p}-i\beta{\bf {\hat  r}}W(r)) +\beta M+V(r)
\end{equation}
where ${\bf {\hat r}} = {{\bf r}\over r}$.The vector $(V(r),{\bf{\hat r}}W(r))$ is interpreted as an external electromagnetic field.Due to the matrix $\beta$ accompanying $W$ in the Hamiltonian, 
the interpretation of  the vector $ (V,{\bf {\hat r}} W)$ as an electromagnetic potential is not necessary and in fact plays no role in his calculations.
The resulting radial equation
\begin{equation}
\lbrack -i\rho_2{d\over dr} +\rho_1(W+{\kappa\over r})-E+V+M\rho_3\rbrack\Phi=0
\end{equation}
where $\Phi=\pmatrix{G_{\ell j}(r)\cr F_{\ell j}(r)\cr}$
corresponds to Alhaidari's equation (1) where the
quantum numbers $\ell$ and $j$ are omitted.
\par
 The subsequent application of a unitary transformation and the imposition of the constraint (in our notation)
\begin{equation}
W(r)={1\over S}V(r)-{\kappa\over r}
\end{equation}
with both $V$ and $W$ nonzero and $S$ a constant can only be satisfied for a chosen value of $\kappa$.Otherwise we will have different functions $W$ for different values of $\kappa$.This cannot be since the functions $V(r)$ and $W(r)$
appear in the Hamiltonian. Forgetting this Alhaidari writes results for the Dirac-Rosen-Morse and Dirac-Eckart potentials which cannot be correct since the energy levels obtained would be degenerate in $l,j,m$. In the nonrelativistic Schroedinger equation the radial equation does contain the centrifugal barrier contribution for nonzero values of $\ell$.
\par
Even if one interprets the results as corresponding to $\ell=0$ so that $\kappa= -1$,the unitary transformation is inexplicable since it does not reduce to identity in the nonrelativistic limit.
\par
The subsequent calculations are for the case $V=0$ treated by Castanõs et al[4] earlier.The results need to be corrected since $ \ell=0$ means $\kappa =-1$ and not $\kappa=0$ as stated by Alhaidari.This has the effect of replacing $W$ by $W-{1\over r}$.In fact, in this restricted case one can find suitable values of $W$ for the Morse,Rosen-Morse and Eckart problems,without the need of any unitary transformation.
\par
Finally,even if the $\ell=0$ case can be adjusted to give a reasonable nonrelativistic limit, the validity of the proposed Dirac equation remains unproved.

\end{document}